\documentclass[preprintnumbers,amsmath,amssymbm,prd]{revtex4}
\usepackage{epsfig}
\usepackage{graphicx}
\usepackage{amssymb}

\begin{document}
\title{Ultra-spinning exotic compact objects supporting static massless scalar field configurations}
\author{Shahar Hod}
\affiliation{The Ruppin Academic Center, Emeq Hefer 40250, Israel}
\affiliation{ } \affiliation{The Hadassah Institute, Jerusalem
91010, Israel}
\date{\today}

\begin{abstract}
\ \ \ Horizonless spacetimes describing highly compact exotic
objects with reflecting (instead of absorbing) surfaces have
recently attracted much attention from physicists and mathematicians
as possible quantum-gravity alternatives to canonical classical
black-hole spacetimes. Interestingly, it has recently been proved
that spinning compact objects with angular momenta in the {\it
sub}-critical regime ${\bar a}\equiv J/M^2\leq1$ are characterized
by an {\it infinite} countable set of surface radii,
$\{r_{\text{c}}({\bar a};n)\}^{n=\infty}_{n=1}$, that can support
asymptotically flat static configurations made of massless scalar
fields. In the present paper we study {\it analytically} the
physical properties of {\it ultra}-spinning exotic compact objects
with dimensionless angular momenta in the complementary regime
${\bar a}>1$. It is proved that ultra-spinning reflecting compact
objects with dimensionless angular momenta in the {\it
super}-critical regime $\sqrt{1-[{{m}/{(l+2)}}]^2}\leq|{\bar
a}|^{-1}<1$ are characterized by a {\it finite} discrete family of
surface radii, $\{r_{\text{c}}({\bar
a};n)\}^{n=N_{\text{r}}}_{n=1}$, distributed symmetrically around
$r=M$, that can support spatially regular static configurations of
massless scalar fields (here the integers $\{l,m\}$ are the harmonic
indices of the supported static scalar field modes). Interestingly,
the largest supporting surface radius
$r^{\text{max}}_{\text{c}}({\bar a})\equiv
\text{max}_n\{r_{\text{c}}({\bar a};n)\}$ marks the onset of
superradiant instabilities in the composed
ultra-spinning-exotic-compact-object-massless-scalar-field system.
\end{abstract}
\bigskip
\maketitle


\section{Introduction}

Curved black-hole spacetimes with absorbing event horizons are one
of the most exciting predictions of the classical Einstein field
equations. The physical and mathematical properties of classical
black-hole spacetimes have been extensively explored during the last
five decades \cite{ThWe,Chan}, and it is widely believed that the
recent detection of gravitational waves \cite{GW1,GW2} provides
compelling evidence for the existence of spinning astrophysical
black holes of the Kerr family. Intriguingly, however, the physical
properties of highly compact {\it horizonless} objects have recently
been explored by many physicists (see
\cite{eco1,eco2,eco3,eco4,eco5,eco6,eco7,eco8,eco9,eco10,eco11,eco12,Pan,Hodeco1,Hodeco2,PanCar,Kz,Hernw}
and references therein) in an attempt to determine whether these
exotic curved spacetimes can serve as valid alternatives, possibly
within the framework of a unified quantum theory of gravity, to
canonical black-hole spacetimes.

In a very interesting work, Maggio, Pani, and Ferrari \cite{Pan}
have recently explored the complex resonance spectrum of massless
scalar fields linearly coupled to horizonless spinning exotic
compact objects. The numerical results presented in \cite{Pan} have
explicitly demonstrated the important physical fact that, for given
values $\{l,m\}$ of the scalar field harmonic indices, there is a
critical compactness parameter characterizing the central reflecting
objects, above which the massless scalar fields grow exponentially
in time. This characteristic behavior of the fields in the
horizonless spinning curved spacetimes indicates that the
corresponding exotic objects may become unstable when coupled to
bosonic (integer-spin) fields \cite{Noteuns}. In particular, this
superradiant instability \cite{Zel,PressTeu2,Viln,Frid,Bri} is
attributed to the fact that the characteristic absorbing boundary
conditions of classical black-hole spacetimes have been replaced in
\cite{Pan} by reflecting boundary conditions at the compact surfaces
of the horizonless exotic objects.

The physical properties of {\it marginally}-stable spinning exotic
compact objects were studied analytically in \cite{Hodeco2}. In
particular, it was explicitly proved in \cite{Hodeco2} that
reflecting compact objects with {\it sub}-critical angular momenta
in the regime $0<{\bar a}\equiv J/M^2\leq1$ \cite{Noteun,Notejm} are
characterized by an {\it infinite} countable set of surface radii,
$\{r_{\text{c}}({\bar a};n)\}^{n=\infty}_{n=1}$, which can support
spatially regular static (marginally-stable) configurations made of
massless scalar fields. The ability of spinning compact objects to
support static scalar field configurations is physically interesting
from the point of view of the no-hair theorems discussed in
\cite{Hodrec,Hodrec2,Bha}. In particular, it was proved in
\cite{Hodrec,Hodrec2} that spherically-symmetric (non-spinning)
horizonless reflecting objects, like black holes with absorbing
horizons \cite{Chas,Bek1,Teit}, cannot support spatially regular
nonlinear massless scalar field configurations
\cite{Notemsm,Hodrc,Herkr}.

Interestingly, the parameter space of the composed
spinning-exotic-compact-object-massless-scalar-field system is
divided by the outermost supporting radius,
$r^{\text{max}}_{\text{c}}({\bar a})\equiv
\text{max}_n\{r_{\text{c}}({\bar a};n)\}$, to stable and unstable
configurations. In particular, horizonless reflecting objects whose
surface radii lie in the regime
$r_{\text{c}}>r^{\text{max}}_{\text{c}}({\bar a})$ are stable to
scalar perturbation modes \cite{Pan,Hodeco2}, whereas the ergoregion
of compact enough spinning objects in the physical regime
$r_{\text{c}}<r^{\text{max}}_{\text{c}}({\bar a})$ can trigger
superradiant instabilities in the surrounding bosonic clouds
\cite{Pan,Hodeco2}.

The main goal of the present paper is to explore the physical
properties of exotic {\it ultra}-spinning (${\bar a}>1$) horizonless
compact objects \cite{Noten1,Herpro,Pet,Herndy}. Interestingly, we
shall explicitly prove below that spinning compact objects in the
{\it super}-critical ${\bar a}>1$ regime are characterized by a {\it
finite} discrete family of surface radii, $\{r_{\text{c}}({\bar
a};n)\}^{n=N_{\text{r}}}_{n=1}$ \cite{Noteeb}, that can support the
static (marginally-stable) scalar field configurations. This unique
property of the {\it ultra}-spinning (${\bar a}>1$) reflecting
compact objects should be contrasted with the previously proved fact
\cite{Hodeco2} that sub-critical (${\bar a}<1$) spinning objects are
characterized by an {\it infinite} countable family of surface
radii, $\{r_{\text{c}}({\bar a};n)\}^{n=\infty}_{n=1}$, that can
support spatially regular static scalar field configurations.

Using analytical techniques, we shall determine in this paper the
characteristic critical (largest) surface radius,
$r^{\text{max}}_{\text{c}}({\bar a})\equiv
\text{max}_n\{r_{\text{c}}({\bar a};n)\}$, of the ultra-spinning
reflecting objects that, for given value of the super-critical
rotation parameter ${\bar a}$, marks the boundary between stable and
superradiantly unstable spinning configurations. In particular,
below we shall derive a remarkably compact analytical formula for
the discrete (and finite) family of supporting surface radii which
characterizes exotic near-critical spinning horizonless compact
objects in the physically interesting regime $0<{\bar a}-1\ll1$.

\section{Description of the system}

We consider a spatially regular configuration made of a massless
scalar field $\Psi$ which is linearly coupled to an ultra-spinning
reflecting compact object of radius $r_{\text{c}}$, mass $M$, and
dimensionless angular momentum in the super-critical regime
\begin{equation}\label{Eq1}
{\bar a}\equiv {{J}\over{M^2}}>1\  .
\end{equation}
Following the interesting physical model of the exotic compact
objects discussed by Maggio, Pani, and Ferrari \cite{Pan} (see also
\cite{Hodeco1,Hodeco2,PanCar}), we shall assume that the external
spacetime geometry of the spinning compact object is described by
the Kerr line element
\cite{ThWe,Chan,Noteun,Notebl,Noteman,Notesk,kr1,kr2,kr3}
\begin{eqnarray}\label{Eq2}
ds^2=-{{\Delta}\over{\rho^2}}(dt-a\sin^2\theta
d\phi)^2+{{\rho^2}\over{\Delta}}dr^2+\rho^2
d\theta^2+{{\sin^2\theta}\over{\rho^2}}\big[a
dt-(r^2+a^2)d\phi\big]^2\ \ \ \text{for}\ \ \ \ r>r_{\text{c}}\  ,
\end{eqnarray}
where the metric functions are given by $\Delta\equiv r^2-2Mr+a^2$
and $\rho^2\equiv r^2+a^2\cos^2\theta$ with $a\equiv M{\bar a}$.

The spatial and temporal behavior of the massless scalar field
configurations in the curved spacetime (\ref{Eq2}) of the spinning
reflecting object is governed by the compact Klein-Gordon wave
equation \cite{Teuk,Stro}
\begin{equation}\label{Eq3}
\nabla^\nu\nabla_{\nu}\Psi=0\  .
\end{equation}
Using the spatial-temporal expression \cite{Teuk,Stro,Notedec}
\begin{equation}\label{Eq4}
\Psi(t,r,\theta,\phi)=\sum_{l,m}e^{im\phi}{S_{lm}}(\theta;a\omega){R_{lm}}(r;M,a,\omega)e^{-i\omega
t}\
\end{equation}
for the linearized massless scalar field, one finds the ordinary
differential equation \cite{Teuk,Stro}
\begin{equation}\label{Eq5}
\Delta{{d}
\over{dr}}\Big(\Delta{{dR_{lm}}\over{dr}}\Big)+\Big\{[\omega(r^2+a^2)-ma]^2
+\Delta(2ma\omega-K_{lm})\Big\}R_{lm}=0\
\end{equation}
for the radial part $R_{lm}(r;M,a,\omega)$ of the massless scalar
eigenfunction. The frequency-dependent eigenvalues $K_{lm}(a\omega)$
of the familiar spheroidal harmonic functions
$S_{lm}(\theta;a\omega)$
\cite{Heun,Fiz1,Teuk,Abram,Stro,Hodasy,Hodpp} are given by the small
frequency $a\omega\ll1$ expression
\begin{equation}\label{Eq6}
K_{lm}-a^2\omega^2=l(l+1)+\sum_{k=1}^{\infty}c_k (a\omega)^{2k}\ ,
\end{equation}
where the explicit functional expression of the coefficients
$\{c_k=c_k(l,m)\}$ is given in \cite{Abram}.

Following the interesting physical models discussed in
\cite{Pan,Hodeco1,Hodeco2,PanCar} for horizonless curved spacetimes,
we shall assume that the scalar fields vanish on the compact
reflecting surfaces of the central exotic compact objects
\cite{Noteby}:
\begin{equation}\label{Eq7}
R(r=r_{\text{c}})=0\  .
\end{equation}
In addition, we consider asymptotically flat linearized scalar field
configurations which are characterized by asymptotically decaying
radial eigenfunctions:
\begin{equation}\label{Eq8}
R(r\to\infty)\to 0\  .
\end{equation}

\section{The resonance condition of the composed ultra-spinning-exotic-compact-
object-massless-scalar-field configurations}

In the present section we shall derive, for a given set of the
dimensionless physical parameters $\{r_{\text{c}}/M,{\bar a},l,m\}$,
the characteristic resonance condition for the existence of
ultra-spinning reflecting exotic horizonless objects that support
spatially regular {\it static} (marginally-stable) linearized scalar
field configurations.

Substituting into the radial equation (\ref{Eq5}) the characteristic
relation
\begin{equation}\label{Eq9}
\omega=0\
\end{equation}
for the static scalar field configurations, one obtains the ordinary
differential equation \cite{Hodeco2,Noteklm0}
\begin{equation}\label{Eq10}
x(1-x){{d^2F}\over{dx^2}}+\{(1-\gamma)-[1+2(l+1)-\gamma]x\}{{dF}\over{dx}}-
[(l+1)^2-\gamma(l+1)]F=0\  ,
\end{equation}
where
\begin{equation}\label{Eq11}
R(x)=x^{-\gamma/2}(1-x)^{l+1}F(x)\  ,
\end{equation}
\begin{equation}\label{Eq12}
x\equiv {{r-M(1+i\sqrt{{\bar a}^2-1})}\over{r-M(1-i\sqrt{{\bar
a}^2-1})}}\  ,
\end{equation}
and
\begin{equation}\label{Eq13}
\gamma\equiv{{m}\over{\sqrt{1-{\bar a}^{-2}}}}\ .
\end{equation}

The physically acceptable solution of the characteristic radial
scalar equation (\ref{Eq10}) which respects the asymptotic boundary
condition (\ref{Eq8}) is given by
\cite{Abram,Morse,Hodeco2,Noteasym}
\begin{equation}\label{Eq14}
R(x)=A\cdot
x^{-\gamma/2}(1-x)^{l+1}{_2F_1}(l+1-\gamma,l+1;2l+2;1-x)\ ,
\end{equation}
where $A$ is a normalization constant and $_2F_1(a,b;c;z)$ is the
hypergeometric function.

Substituting the radial solution (\ref{Eq14}) into the
characteristic inner boundary condition (\ref{Eq7}) at the surface
of the compact reflecting object, one obtains the remarkably compact
resonance condition
\begin{equation}\label{Eq15}
{_2F_1}(l+1-\gamma,l+1;2l+2;1-x_{\text{c}})=0\
\end{equation}
for the composed
ultra-spinning-exotic-compact-object-massless-scalar-field
configurations.

As we shall show below, the resonance equation (\ref{Eq15})
determines the {\it discrete} set of surface radii
$\{r_{\text{c}}=r_{\text{c}}({\bar a},l,m;n)\}$ which characterize
the unique family of ultra-spinning exotic compact objects that can
support the static spatially regular massless scalar field
configurations.

\section{Generic properties of the composed ultra-spinning-exotic-compact-
object-massless-scalar-field configurations}

In the present section we shall discuss two important features of
the discrete resonance spectrum $\{r_{\text{c}}({\bar a},l,m;n)\}$
of surface radii that characterize the composed
ultra-spinning-exotic-compact-object-massless-scalar-field
configurations: (1) the distribution of the supporting radii, and
(2) the (finite) number of supporting radii.

\subsection{The resonance spectrum of surface radii is distributed symmetrically
around $r=M$}

Interestingly, we shall now prove that the discrete set of
supporting radii $\{r_{\text{c}}({\bar a},l,m;n)\}$, which stems
from the characteristic resonance equation (\ref{Eq15}), is
distributed {\it symmetrically} around $r=M$. To this end, it is
convenient to define the dimensionless symmetrical radial coordinate
\begin{equation}\label{Eq16}
z\equiv {{r-M}\over{M}}\  ,
\end{equation}
in terms of which the resonance equation (\ref{Eq15}) can be written
in the form \cite{Notexz}
\begin{equation}\label{Eq17}
{_2F_1}\Big(l+1-{{m}\over{\sqrt{1-{\bar
a}^{-2}}}},l+1;2l+2;{{2i\sqrt{{\bar
a}^2-1}}\over{z_{\text{c}}+i\sqrt{{\bar a}^2-1}}}\Big)=0\  .
\end{equation}
Using the characteristic identity (see Eq. 15.3.15 of \cite{Abram})
\begin{equation}\label{Eq18}
{_2F_1}(a,b;2b;z)=(1-z)^{-a/2}{_2F_1}({1\over2}a,b-{1\over2}a;b+{1\over2};{{z^2}\over{4z-4}})\
\end{equation}
of the hypergeometric function, one can express the resonance
condition (\ref{Eq17}) in the symmetrical form
\begin{equation}\label{Eq19}
{_2F_1}\Big[{1\over2}\Big(l+1-{{m}\over{\sqrt{1-{\bar
a}^{-2}}}}\Big),{1\over2}\Big(l+1+{{m}\over{\sqrt{1-{\bar
a}^{-2}}}}\Big);l+{3\over2};{{{\bar a}^2-1}\over{{\bar
a}^2-1+z^2_{\text{c}}}}\Big]=0\ .
\end{equation}
The resonance equation (\ref{Eq19}) is obviously invariant under the
reflection symmetry $z_{\text{c}}\to -z_{\text{c}}$. We have
therefore proved that if the dimensionless surface radius
$z_{\text{c}}$ is a solution of the characteristic resonance
equation (\ref{Eq19}), then $-z_{\text{c}}$ is also a valid
resonance.

In addition, it is interesting to stress the fact that, for the
static ($\omega=0$) scalar field modes, the radial scalar equation
(\ref{Eq5}) is invariant under the reflection symmetries $a\to -a$
and $m\to -m$ \cite{Notesmm}. One therefore deduces that if the
dimensionless surface radius $z_{\text{c}}$ characterizes an
ultra-spinning exotic compact object with $ma>0$ that can support a
spatially regular static (marginally-stable) scalar field
configuration with harmonic indices $\{l,m\}$, then the same
supporting radius also characterizes an ultra-spinning exotic
compact object with $ma<0$ that can support the same static scalar
field configuration.

Taking cognizance of the three reflection symmetries,
$z_{\text{c}}\to -z_{\text{c}}$, $a\to -a$, and $m\to -m$, which
characterize the composed
ultra-spinning-exotic-compact-object-massless-scalar-field system,
we shall henceforth assume, without loss of generality, that
\begin{equation}\label{Eq20}
a>0\ \ \ \ ; \ \ \ \ m>0\ \ \ \ ; \ \ \ \ z_{\text{c}}\geq0\ .
\end{equation}

\subsection{The number of discrete supporting radii is finite}

As emphasized above, it has recently been proved \cite{Hodeco2} that
exotic compact objects in the {\it sub}-critical regime ${\bar a}<1$
are characterized by an {\it infinite} set of surface radii,
$\{r_{\text{c}}({\bar a};n)\}^{n=\infty}_{n=1}$, that can support
static (marginally-stable) massless scalar field configurations.

On the other hand, we shall now show that {\it super}-critical
(${\bar a}>1$) compact reflecting objects are characterized by a
{\it finite} set of surface radii that can support the static
massless scalar field configurations. In particular, one finds that,
for positive integer values of the dimensionless physical parameter
$N({\bar a},l,m)\equiv \gamma-(l+1)$, the resonance equation
(\ref{Eq15}), which determines the characteristic spectrum of
supporting radii of the {\it ultra}-spinning exotic compact objects,
is a polynomial equation of degree $N$. Thus, in this case there is
a {\it finite} number $N$ of complex solutions $\{x_{\text{c}}({\bar
a},l,m;n)\}^{n=N}_{n=1}$ to the resonance condition (\ref{Eq15})
which in turn, using the relation (\ref{Eq12}), yield a finite
discrete spectrum $\{r_{\text{c}}({\bar a},l,m;n)\}^{n=N}_{n=1}$ of
supporting surface radii.

In addition, solving numerically the resonance equation (\ref{Eq15})
we find that, for positive non-integer values of the physical
parameter $N$, the number of discrete surface radii that can support
the static (marginally-stable) scalar field configurations is given
by (see Tables \ref{Table1} and \ref{Table2} below) $\left
\lfloor{N}\right \rfloor$ for even values of $\left \lfloor{N}\right
\rfloor$ and by $\left \lfloor{N}\right \rfloor+1$ for odd values of
$\left \lfloor{N}\right \rfloor$ \cite{Notefl}.

To summarize, the ({\it finite}) number $N_{\text{r}}({\bar a},l,m)$
of discrete supporting radii that characterize the composed
ultra-spinning-exotic-compact-object-massless-scalar-field
configurations is given by the simple relations
\begin{equation}\label{Eq21}
N_{\text{r}}=
\begin{cases}
\gamma-(l+1) & \text{ if }\ \ \ \gamma-(l+1)\ \text{is a positive integer}\ ; \\
\lfloor{\gamma-(l+1)}\rfloor & \text{ if }\ \ \ \lfloor{\gamma-(l+1)}\rfloor\  \text{is a positive even integer}\ ; \\
\lfloor{\gamma-(l+1)}\rfloor+1 & \text{ if }\ \ \
\lfloor{\gamma-(l+1)}\rfloor\  \text{is a positive odd integer}\ .
\end{cases}
\end{equation}
[It is important to emphasize that cases 2 and 3 in (\ref{Eq21})
refer to {\it non}-integer values of the dimensionless composed
parameter $\gamma-(l+1)$].

\section{The regime of existence of the composed ultra-spinning-exotic-compact-
object-massless-scalar-field configurations}

In the present section we shall derive an upper bound on the
characteristic surface radii $\{r_{\text{c}}({\bar
a},l,m;n)\}^{n=N_{\text{r}}}_{n=1}$ which characterize the
ultra-spinning exotic compact objects that can support the static
(marginally-stable) configurations of the massless scalar fields.

Substituting the scalar function
\begin{equation}\label{Eq22}
\Phi(r)\equiv \Delta^{1/2}\cdot R(r)\
\end{equation}
into the characteristic radial equation (\ref{Eq5}), one obtains the
ordinary differential equation
\begin{equation}\label{Eq23}
\Delta^2{{d^2\Phi}\over{dr^2}}+\Big[(ma)^2-l(l+1)\cdot\Delta-(a^2-M^2)\Big]\Phi=0\
\end{equation}
for the static ($\omega=0$) scalar configurations.

Using the characteristic boundary conditions (\ref{Eq7}) and
(\ref{Eq8}) of the spatially regular linearized scalar field
configurations, which are supported in the asymptotically flat
curved spacetime (\ref{Eq2}) of the exotic ultra-spinning reflecting
compact object, one deduces that the radial scalar eigenfunction
$\Phi(r)$ must have (at least) one extremum point,
$r=r_{\text{peak}}$, in the interval
\begin{equation}\label{Eq24}
r_{\text{peak}}\in (r_{\text{c}},\infty)\  .
\end{equation}
In particular, the simple functional relations
\begin{equation}\label{Eq25}
\{\Phi\neq 0\ \ \ ; \ \ \ {{d\Phi}\over{dr}}=0\ \ \ ; \ \ \
\Phi\cdot{{d^2\Phi}\over{dr^2}}<0\}\ \ \ \ \text{for}\ \ \ \
r=r_{\text{peak}}\
\end{equation}
characterize the spatial behavior of the radial scalar eigenfunction
at this extremum point.

Taking cognizance of Eqs. (\ref{Eq23}) and (\ref{Eq25}), one finds
the simple relation
\begin{equation}\label{Eq26}
(ma)^2-l(l+1)\cdot\Delta(r_{\text{peak}})-(a^2-M^2)>0\  .
\end{equation}
The characteristic inequality (\ref{Eq26}) implies that
$r_{\text{peak}}$ is bounded by the relations
\begin{equation}\label{Eq27}
r_-<r_{\text{peak}}<r_+\  ,
\end{equation}
where
\begin{equation}\label{Eq28}
r_{\pm}=M\pm\sqrt{M^2-{{a^2[1+l(l+1)-m^2]-M^2}\over{l(l+1)}}}\ .
\end{equation}
Using Eqs. (\ref{Eq16}), (\ref{Eq24}), (\ref{Eq27}), and
(\ref{Eq28}), one deduces that the composed
ultra-spinning-exotic-compact-object-massless-scalar-field
configurations are characterized by the simple dimensionless upper
bound
\begin{equation}\label{Eq29}
|z_{\text{c}}|<\sqrt{1-{{{\bar a}^2[1+l(l+1)-m^2]-1}\over{l(l+1)}}}\
.
\end{equation}
In particular, from the requirement ${\bar a}^2[1+l(l+1)-m^2]-1\leq
l(l+1)$ [see the r.h.s of (\ref{Eq29})], one finds that the static
(marginally-stable) massless scalar field configurations in the
curved spacetimes of the ultra-spinning (${\bar a}>1$) exotic
compact objects are characterized by the compact inequalities
\begin{equation}\label{Eq30}
\sqrt{1-{{m^2}\over{1+l(l+1)}}}<|{\bar a}|^{-1}<1\  .
\end{equation}

Interestingly, a stronger upper bound on the dimensionless angular
momentum parameter ${\bar a}$, which characterizes the unique family
of ultra-spinning exotic compact objects that can support the
spatially regular static (marginally-stable) massless scalar field
configurations, can be obtained from the observations that [see Eq.
(\ref{Eq15})] \cite{Notenao}
\begin{equation}\label{Eq31}
{_2F_1}[l+1-\gamma,l+1;2l+2;1-x(r)]\neq0\ \ \ \ \text{for}\ \ \ \
\{r\in \mathbb{R}\ \ \ \text{and}\ \ \ -1<l+1-\gamma<2l+3\}\
\end{equation}
and
\begin{equation}\label{Eq32}
{_2F_1}(-1,l+1;2l+2;2)={_2F_1}(2l+3,l+1;2l+2;2)=0\  .
\end{equation}
From Eqs. (\ref{Eq13}), (\ref{Eq31}), and (\ref{Eq32}), one deduces
that the composed
ultra-spinning-exotic-compact-object-massless-scalar-field
configurations exist in the dimensionless physical regime
\begin{equation}\label{Eq33}
\sqrt{1-\Big({{m}\over{l+2}}\Big)^2}\leq|{\bar a}|^{-1}<1\  ,
\end{equation}
where the equality sign in (\ref{Eq33}) corresponds to exotic
ultra-spinning objects with $r_{\text{c}}=M$ [or, equivalently,
$1-x_{\text{c}}=2$ and $z_{\text{c}}=0$, see Eqs. (\ref{Eq12}) and
(\ref{Eq16})].

\section{The resonance spectrum of the composed ultra-spinning-exotic-compact-
object-massless-scalar-field configurations}

As mentioned above, the {\it infinite} countable spectrum of
supporting surface radii $\{r_{\text{c}}({\bar
a},l,m;n)\}^{n=\infty}_{n=1}$ which characterizes the composed
spinning-exotic-compact-object-massless-scalar-field configurations
in the {\it sub}-critical regime ${\bar a}<1$ has been determined in
\cite{Hodeco2}. In the present section we shall explicitly show that
{\it ultra}-spinning exotic compact objects in the complementary
regime ${\bar a}>1$ of super-critical angular momenta are
characterized by a {\it finite} [see Eq. (\ref{Eq21})] discrete set
$\{r_{\text{c}}({\bar a},l,m;n)\}^{n=N_{\text{r}}}_{n=1}$ of surface
radii that can support the asymptotically flat static scalar field
configurations.

The compact resonance equation (\ref{Eq15}) can be solved
numerically, for a given set $\{{\bar a},l,m\}$ of the dimensionless
physical parameters that characterize the composed
compact-object-scalar-field system, to yield the discrete resonant
spectrum $\{r_{\text{c}}({\bar a},l,m;n)\}^{n=N_{\text{r}}}_{n=1}$
of supporting radii. In Table \ref{Table1} we present, for various
super-critical values of the dimensionless angular momentum
parameter ${\bar a}$, the smallest and largest dimensionless surface
radii $\{z^{\text{min}}_{\text{c}}({\bar
a},l,m),z^{\text{max}}_{\text{c}}({\bar a},l,m)\}$ of the
ultra-spinning exotic compact objects that can support the static
spatially regular configurations of the massless scalar fields
\cite{Notezp}. We also present in Table \ref{Table1} the ({\it
finite}) number $N_{\text{r}}({\bar a},l,m)$ [see Eq. (\ref{Eq21})]
of these unique supporting surface radii \cite{Noteodd}.


The data presented in Table \ref{Table1} demonstrate the fact that,
for given integer values $\{l,m\}$ of the angular harmonic indices
of the static (marginally-stable) massless scalar fields, the
dimensionless supporting radius $z^{\text{max}}_{\text{c}}({\bar
a})$ of the ultra-spinning exotic compact objects is a monotonically
decreasing function of the dimensionless physical parameter ${\bar
a}$. As a consistency check, it is worth noting that the numerically
computed values $\{z^{\text{max}}_{\text{c}}({\bar a})\}$ of the
characteristic surface radii of the ultra-spinning reflecting
compact objects, as displayed in Table \ref{Table1}, conform to the
analytically derived upper bound (\ref{Eq29}).

We would like to emphasize again that, for a given set of the
physical parameters $\{{\bar a},l,m\}$, the critical supporting
radius $r^{\text{max}}_{\text{c}}({\bar a})$ marks the boundary
between stable and superradiantly unstable composed
ultra-spinning-exotic-compact-object-massless-scalar-field
configurations. In particular, the numerical results presented in
the interesting work of Maggio, Pani, and Ferrari \cite{Pan}
indicate that ultra-spinning reflecting compact objects which are
characterized by the inequality
$r_{\text{c}}<r^{\text{max}}_{\text{c}}({\bar a})$ are
superradiantly unstable to massless scalar perturbation modes,
whereas ultra-spinning exotic compact objects which are
characterized by the relation
$r_{\text{c}}>r^{\text{max}}_{\text{c}}({\bar a})$ are stable.

\begin{table}[htbp]
\centering
\begin{tabular}{|c|c|c|c|c|}
\hline $\ \ \sqrt{1-{\bar a}^{-2}}\ \ $\ \ &\ \ ${\bar a}\ \ $\ \ &
\ \ $\#$\ \text{of resonances} \ \ & \ $\ \
z^{\text{min}}_{\text{c}}({\bar a})\ \ $ \ \ & \
$\ \ z^{\text{max}}_{\text{c}}({\bar a})\ \ $\ \ \\
\hline
\ \ $1/3$\ \ \ &\ \ \ $1.0607$\ \ \ \ & \ \ $1$\ \ &\ \ $0$\ \ &\ \ $0$\ \ \\
\ \ $0.3$\ \ \ &\ \ \ $1.0483$\ \ \ \ & \ \ $2$\ \ &\ \ $0.05488$\ \ &\ \ $0.05488$\ \ \\
\ \ $0.25$\ \ \ &\ \ \ $1.0328$\ \ \ \ & \ \ $2$\ \ &\ \ $0.11547$\ \ &\ \ $0.11547$\ \ \\
\ \ $0.2$\ \ \ &\ \ \ $1.0206$\ \ \ \ & \ \ $3$\ \ &\ \ $0$\ \ &\ \ $0.15811$\ \ \\
\ \ $0.15$\ \ \ &\ \ \ $1.0114$\ \ \ \ & \ \ $4$\ \ &\ \ $0.06455$\ \ &\ \ $0.18788$\ \ \\
\ \ $0.1$\ \ \ &\ \ \ $1.0050$\ \ \ \ & \ \ $8$\ \ &\ \ $0.01608$\ \ &\ \ $0.20760$\ \ \\
\hline
\end{tabular}
\caption{Marginally-stable ultra-spinning (${\bar a}>1$) reflecting
compact objects. We present, for various super-critical values of
the dimensionless physical parameter ${\bar a}$, the smallest and
largest dimensionless radii, $\{z^{\text{min}}_{\text{c}}({\bar
a},l,m),z^{\text{max}}_{\text{c}}({\bar a},l,m)\}$ [see Eq.
(\ref{Eq16})], of the ultra-spinning exotic compact objects that can
support the static (marginally-stable) massless scalar field
configurations \cite{Notezp}. Also presented is the {\it finite}
number [see Eq. (\ref{Eq21})] of these unique supporting surface
radii. The data presented is for the case $l=m=1$. The critical
supporting radii $\{z^{\text{max}}_{\text{c}}({\bar a})\}$, which
characterize the marginally-stable ultra-spinning reflecting compact
objects, are found to be a monotonically decreasing function of the
dimensionless angular momentum parameter ${\bar a}$. As a
consistency check we note that the supporting radii of the
ultra-spinning exotic compact objects conform to the analytically
derived upper bound (\ref{Eq29}).} \label{Table1}
\end{table}

In Table \ref{Table2} we present, for various equatorial ($l=m$)
modes of the supported static scalar fields, the smallest and
largest dimensionless surface radii
$\{z^{\text{min}}_{\text{c}}({\bar
a},l,m),z^{\text{max}}_{\text{c}}({\bar a},l,m)\}$ of the supporting
marginally-stable ultra-spinning exotic compact objects
\cite{Notezp}. Also displayed is the ({\it finite}) number [see Eq.
(\ref{Eq21})] of these unique supporting surface radii
\cite{Noteodd}. The data presented in Table \ref{Table2} reveal the
fact that, for a given value of the dimensionless physical parameter
${\bar a}$, the critical (largest) supporting radius
$z^{\text{max}}_{\text{c}}(l)$ of the reflecting exotic compact
objects is a monotonically increasing function of the harmonic index
$l$ which characterizes the static massless scalar field mode. It is
worth noting that the numerically computed surface radii
$z^{\text{max}}_{\text{c}}(l)$ of the ultra-spinning
marginally-stable exotic compact objects, as presented in Table
\ref{Table2}, conform to the analytically derived upper bound
(\ref{Eq29}).

\begin{table}[htbp]
\centering
\begin{tabular}{|c|c|c|c|}
\hline $\ \ \ l\ \ \ $\ \ & \ \ $\#$\ \text{of resonances} \ \ & \
$\ \ z^{\text{min}}_{\text{c}}(l)\ \ $ \ \ & \
$\ \ z^{\text{max}}_{\text{c}}(l)\ \ $\ \ \\
\hline
\ \ $1$\ \ \ &\ \ $2$\ \ &\ \ $0.11547$\ \ &\ \ $0.11547$\ \ \\
\ \ $2$\ \ \ &\ \ $5$\ \ &\ \ $0$\ \ &\ \ $0.28705$\ \ \\
\ \ $3$\ \ \ &\ \ $8$\ \ &\ \ $0.03553$\ \ &\ \ $0.38489$\ \ \\
\ \ $4$\ \ \ &\ \ $11$\ \ &\ \ $0$\ \ &\ \ $0.45263$\ \ \\
\ \ $5$\ \ \ &\ \ $14$\ \ &\ \ $0.02113$\ \ &\ \ $0.50342$\ \ \\
\ \ $6$\ \ \ &\ \ $17$\ \ &\ \ $0$\ \ &\ \ $0.54338$\ \ \\
\hline
\end{tabular}
\caption{Marginally-stable ultra-spinning (${\bar a}>1$) reflecting
compact objects. We present, for various equatorial ($l=m$) modes of
the supported scalar fields, the smallest and largest dimensionless
surface radii $\{z^{\text{min}}_{\text{c}}({\bar
a},l,m),z^{\text{max}}_{\text{c}}({\bar a},l,m)\}$ [see Eq.
(\ref{Eq16})] of the ultra-spinning exotic compact objects that can
support the spatially regular static (marginally-stable) massless
scalar field configurations \cite{Notezp}. We also present the {\it
finite} number of these unique supporting radii. The data presented
is for the case $\sqrt{1-{\bar a}^{-2}}=1/4$. The critical surface
radii $\{z^{\text{max}}_{\text{c}}(l)\}$, which characterize the
marginally-stable ultra-spinning exotic compact objects, are found
to be a monotonically increasing function of the dimensionless
harmonic index $l$ of the supported static scalar field
configurations.} \label{Table2}
\end{table}

\section{The resonance spectrum of near-critical ultra-spinning exotic compact objects}

\subsection{An analytical treatment}

Interestingly, as we shall explicitly show in the present section,
the compact resonance equation (\ref{Eq15}), which determines the
discrete family $\{x_{\text{c}}({\bar a},l,m;n)\}$ of dimensionless
surface radii that characterize the marginally-stable ultra-spinning
exotic compact objects, is amenable to an {\it analytical} treatment
in the physically interesting regime
\begin{equation}\label{Eq34}
0<{\bar a}-1\ll1\
\end{equation}
of {\it near-critical} horizonless spinning configurations.

In particular, in the near-critical regime
\begin{equation}\label{Eq35}
{{m}\over{\sqrt{1-{\bar a}^{-2}}}}\gg l\
\end{equation}
one may use the large-$|b|$ asymptotic expansion \cite{Itn}
\begin{equation}\label{Eq36}
{_2F_1}(a,b;c;z)={{\Gamma(c)}\over{\Gamma(c-a)}}(-bz)^{-a}[1+O(|bz|^{-1})]
+{{\Gamma(c)}\over{\Gamma(a)}}(bz)^{a-c}(1-z)^{c-a-b}[1+O(|bz|^{-1})]
\end{equation}
of the hypergeometric function in order to express the resonance
condition (\ref{Eq15}) in the remarkably compact form \cite{Notefp1}
\begin{equation}\label{Eq37}
x^{m/\sqrt{1-{\bar a}^{-2}}}=(-1)^{-l}\ \ \ \ \text{for}\ \ \ \
\sqrt{1-{\bar a}^{-2}}\ll {{m}\over{l}}\  .
\end{equation}
From the asymptotic relation (\ref{Eq37}) one finds the set of
complex solutions \cite{Notenn}
\begin{equation}\label{Eq38}
x_{\text{c}}(n)=e^{-i\pi(l+2n)\sqrt{1-{\bar a}^{-2}}/m}\ \ \ ; \ \ \
n\in\mathbb{Z}\
\end{equation}
which, taking cognizance of Eqs. (\ref{Eq12}) and (\ref{Eq16}),
yields the discrete real resonance spectrum
\cite{Notexz,Notefp1,Notefp2}
\begin{equation}\label{Eq39}
z_{\text{c}}(n)=\sqrt{{\bar
a}^{2}-1}\cdot\cot\Big[{{\pi(l+2n)\sqrt{1-{\bar
a}^{-2}}}\over{2m}}\Big]\ \ \ \ \text{for}\ \ \ \
\big\{\sqrt{1-{\bar a}^{-2}}\ll {{m}\over{l}}\ \ \ \text{and}\ \ \
\pi(l+2n)\gg1\big\}\
\end{equation}
for the dimensionless surface radii which characterize the
near-critical (${\bar a}\gtrsim1$) exotic compact objects that can
support the static massless scalar field configurations.
Interestingly, the analytically derived resonance formula
(\ref{Eq39}) can be further simplified in the
${{\pi(l+2n)\sqrt{1-{\bar a}^{-2}}}/{2m}}\ll1$ regime, in which case
one finds the remarkably compact expression \cite{Noteapr,Notefp2}
\begin{equation}\label{Eq40}
z_{\text{c}}(n)={{2m{\bar a}}\over{\pi(l+2n)}}\ \ \ \ \text{for}\ \
\ \ 1\ll \pi(l+2n)\ll {{2m}\over{\sqrt{1-{\bar a}^{-2}}}}\
\end{equation}
for the characteristic radii of the ultra-spinning exotic compact
objects that can support the static (marginally-stable)
configurations of the massless scalar fields.

\subsection{Numerical confirmation}

It is of physical interest to verify the accuracy of the
approximated (analytically derived) resonance spectrum (\ref{Eq39})
for the surface radii of the near-critical ($0<{\bar a}-1\ll1$)
ultra-spinning exotic compact objects that can support the spatially
regular static (marginally-stable) configurations of the massless
scalar fields. In Table \ref{Table3} we present the dimensionless
discrete surface radii $z^{\text{analytical}}_{\text{c}}(n)$ of the
supporting near-critical ultra-spinning exotic reflecting objects as
obtained from the analytically derived resonance spectrum
(\ref{Eq39}). We also present in Table \ref{Table3} the
corresponding surface radii $z^{\text{numerical}}_{\text{c}}(n)$ of
the ultra-spinning exotic compact objects as computed numerically
from the exact characteristic resonance equation (\ref{Eq15}).

The data presented in Table \ref{Table3} nicely demonstrate the
important fact that there is a good agreement
between the approximated surface radii
$\{z^{\text{analytical}}_{\text{c}}(n)\}$ of the ultra-spinning
exotic compact objects that can support the static massless scalar
field configurations [as calculated from the compact analytically
derived resonance formula (\ref{Eq39})] and the corresponding exact
surface radii $\{z^{\text{numerical}}_{\text{c}}(n)\}$ of the
reflecting compact objects [as determined numerically directly from
the resonance equation (\ref{Eq15})].

\begin{table}[htbp]
\centering
\begin{tabular}{|c|c|c|c|c|c|}
\hline \text{Formula} & \ $\ z_{\text{c}}(n=1)\ $\ \ & \ $\
z_{\text{c}}(n=2)\ $\ \ & \
$\ z_{\text{c}}(n=3)\ $\ \ & \ $\ z_{\text{c}}(n=4)\ $\ \ & \ $\ z_{\text{c}}(n=5)\ $\ \ \\
\hline \ {\text{Analytical}}\ [Eq. (\ref{Eq39})]\ \ &\ $0.21206$\ \
&\ $0.12707$\ \
&\ $0.09058$\ \ &\ $0.07027$\ \ &\ $0.05730$\ \ \\
\ {\text{Numerical}}\ [Eq. (\ref{Eq15})]\ \ &\ $0.22240$\ \ &\
$0.12919$\ \ &\ $0.09135$\ \ &\ $0.07062$\ \
&\ $0.05818$\ \ \\
\hline
\end{tabular}
\caption{Near-critical ultra-spinning (${\bar a}\gtrsim1$) exotic
compact objects. Displayed are the analytically calculated discrete
surface radii $\{z^{\text{analytical}}_{\text{c}}(n)\}$ which
characterize the ultra-spinning exotic compact objects that can
support the static (marginally-stable) spatially regular
configurations of the massless scalar fields. Also displayed are the
corresponding supporting radii
$\{z^{\text{numerical}}_{\text{c}}(n)\}$ of the near-critical exotic
compact objects as obtained numerically directly from the
characteristic resonance equation (\ref{Eq15}). The data presented
is for static massless scalar field modes with $l=m=1$ linearly
coupled to near-critical ultra-spinning exotic compact objects with
$\sqrt{1-{\bar a}^{-2}}=10^{-2}$. The displayed data reveal a
remarkably good agreement between the exact characteristic surface
radii $\{z^{\text{numerical}}_{\text{c}}(n)\}$ of the ultra-spinning
exotic compact objects [as determined numerically from the resonance
condition (\ref{Eq15})] and the corresponding approximated radii
$\{z^{\text{analytical}}_{\text{c}}(n)\}$ of the near-critical
ultra-spinning compact objects [as calculated analytically from the
compact resonance formula (\ref{Eq39})].} \label{Table3}
\end{table}

\section{The resonance spectrum of ultra-spinning exotic compact objects with $r_{\text{c}}=M$}

Interestingly, as we shall now prove, the characteristic resonance
equation (\ref{Eq15}) [or, equivalently, the symmetrical form
(\ref{Eq19}) of the resonance condition] can also be solved {\it
analytically} for the dimensionless angular momentum parameter
${\bar a}$ in the physically interesting case of horizonless
ultra-spinning exotic compact objects of mass $M$ whose compact
reflecting surfaces coincide with the corresponding horizon radius
$r_{\text{c}}=M$ of extremal Kerr black holes with the same mass
parameter.

Substituting $r_{\text{c}}=M$ [which corresponds to
$z_{\text{c}}=0$, see Eq. (\ref{Eq16})] into the resonance condition
(\ref{Eq19}), and using the characteristic identity (see Eq. 15.1.20
of \cite{Abram})
\begin{equation}\label{Eq41}
{_2F_1}(a,b;c;1)={{\Gamma(c)\Gamma(c-a-b)}\over{\Gamma(c-a)\Gamma(c-b)}}\
\end{equation}
of the hypergeometric function, one finds the compact resonance
equation
\begin{equation}\label{Eq42}
{{\Gamma\big(l+{{3}\over{2}}\big)\Gamma\big({{1}\over{2}}\big)}
\over{\Gamma\big({1\over2}l+1+{{m}\over{2\sqrt{1-{\bar
a}^{-2}}}}\big)\Gamma\big({1\over2}l+1-{{m}\over{2\sqrt{1-{\bar
a}^{-2}}}}\big)}}=0\ \ \ \ \text{for}\ \ \ \ r_{\text{c}}=M\  .
\end{equation}
Using the well known pole structure of the Gamma functions [namely,
$1/\Gamma(-n)=0$ for $n=0,1,2,...$ \cite{Noteag}], one obtains from
(\ref{Eq42}) the remarkably simple discrete resonance spectrum
\begin{equation}\label{Eq43}
\sqrt{1-{\bar a}^{-2}}={{m}\over{l+2+2n}}\ \ \ ; \ \ \ n=0,1,2,...\
\end{equation}
for the ultra-spinning (${\bar a}>1$) exotic compact objects whose
reflecting surfaces coincide with the corresponding horizon radius
$r_{\text{c}}=M$ of extremal (${\bar a}=1$) Kerr black holes
\cite{Noternm}. Interestingly, one finds that the ultra-spinning
exotic compact objects described by the resonance formula
(\ref{Eq43}) with $n=0$ saturate the previously derived bound
(\ref{Eq33}).

\section{Summary and Discussion}

The physical and mathematical properties of horizonless highly
compact exotic reflecting objects have recently been studied by some
physicists (see
\cite{eco1,eco2,eco3,eco4,eco5,eco6,eco7,eco8,eco9,eco10,eco11,eco12,Pan,Hodeco1,Hodeco2,PanCar,Kz,Hernw}
and references therein). The main motivation behind these diverse
studies has been to examine the intriguing possibility that these
exotic horizonless objects may serve as quantum-gravity alternatives
to classical black-hole spacetimes.

Interestingly, Maggio, Pani, and Ferrari \cite{Pan} have recently
provided compelling evidence that sub-critical (${\bar a}\equiv
J/M^2<1$) horizonless spinning spacetimes, in which the
characteristic absorbing boundary conditions of classical black-hole
spacetimes have been replaced by reflective boundary conditions at
the surfaces of the exotic compact objects, may become
superradiantly unstable \cite{Zel,PressTeu2,Viln,Frid,Bri} when
linearly coupled to massless scalar (bosonic) field modes
\cite{Noteuns}. In particular, it has been explicitly proved in
\cite{Hodeco2} that, in the sub-critical regime ${\bar a}<1$ of the
spinning reflecting objects and for given harmonic indices $\{l,m\}$
of the massless scalar field, there exists an {\it infinite}
countable set of surface radii, $\{r_{\text{c}}({\bar
a},l,m;n)\}^{n=\infty}_{n=1}$, which can support spatially regular
static (marginally-stable) massless scalar field configurations.

In the present paper we have explored the physical and mathematical
properties of marginally-stable composed
ultra-spinning-exotic-compact-object-massless-scalar-field
configurations which are characterized by super-critical (${\bar
a}>1$) dimensionless rotation parameters. The following are the main
results derived in this paper and their physical implications:

(1) It has been explicitly proved that, for given dimensionless
physical parameters $\{{\bar a},l,m\}$, the unique discrete family
$\{r_{\text{c}}({\bar a},l,m;n)\}$ of surface radii that
characterize the ultra-spinning (${\bar a}>1$) exotic compact
objects that can support the static (marginally-stable) massless
scalar field configurations is determined by the resonance condition
[see Eqs. (\ref{Eq1}), (\ref{Eq16}), and (\ref{Eq19})]
\begin{equation}\label{Eq44}
{_2F_1}\Big[{1\over2}\Big(l+1-{{ma}\over{\sqrt{a^2-M^2}}}\Big),
{1\over2}\Big(l+1+{{ma}\over{\sqrt{a^2-M^2}}}\Big);l+{3\over2};{{a^2-M^2}
\over{a^2-M^2+(r_{\text{c}}-M)^2}}\Big]=0\  .
\end{equation}

(2) We have shown that the composed
ultra-spinning-exotic-compact-object-massless-scalar-field
configurations, as determined by the resonance condition
(\ref{Eq44}), are restricted to the physical regime [see Eq.
(\ref{Eq33})]
\begin{equation}\label{Eq45}
\sqrt{1-\Big({{m}\over{l+2}}\Big)^2}\leq{{M}\over{a}}<1\
\end{equation}
of the dimensionless super-critical rotation parameter $a/M$. In
addition, it has been proved that, for a given set $\{{\bar
a},l,m\}$ of the dimensionless physical parameters that characterize
the composed compact-object-scalar-field system, the simple relation
[see Eqs. (\ref{Eq16}) and (\ref{Eq29})]
\begin{equation}\label{Eq46}
\Big|{{r_{\text{c}}-M}\over{M}}\Big|<\sqrt{1-{{{\bar
a}^2[1+l(l+1)-m^2]-1}\over{l(l+1)}}}\
\end{equation}
provides an upper bound on the surface radii of the supporting
ultra-spinning exotic compact objects.

(3) It has been pointed out that the analytically derived resonance
condition in its symmetrical form (\ref{Eq44}) reveals the fact
that, for ultra-spinning exotic compact objects, the discrete
resonant spectrum of supporting surface radii is invariant under the
reflection  symmetries
\begin{equation}\label{Eq47}
r_{\text{c}}-M\to -(r_{\text{c}}-M)\ \ \ \ ; \ \ \ \ a\to -a \ \ \ \
; \ \ \ \ m\to -m\  .
\end{equation}
The symmetry transformations (\ref{Eq47}) imply, in particular, that
if $z_{\text{c}}$ [see Eq. (\ref{Eq16})] is a dimensionless
supporting radius of a composed exotic-object-scalar-field system
with dimensionless physical parameters $\{{\bar a},l,m\}$, then: (1)
$-z_{\text{c}}$ is also a valid supporting radius of the same
composed physical system, and (2) $z_{\text{c}}$ is also a valid
supporting radius of a composed exotic-object-scalar-field system
with dimensionless parameters $\{\pm{\bar a},l,\pm m\}$.

(4) It has been shown that, for ultra-spinning exotic compact
objects in the dimensionless physical regime (\ref{Eq45}), the {\it
finite} number $N_{\text{r}}({\bar a},l,m)$ of surface radii that
can support the spatially regular static (marginally-stable) scalar
field configurations is given by [see Eqs. (\ref{Eq13}) and
(\ref{Eq21})] \cite{Notefl,Noternm,NoteNN2}
\begin{equation}\label{Eq48}
N_{\text{r}}=
\begin{cases}
N & \text{ if }\ \ \ N\ \text{is a positive integer}\ ; \\
\lfloor{N}\rfloor & \text{ if }\ \ \ \lfloor{N}\rfloor\  \text{is a positive even integer}\ ; \\
\lfloor{N}\rfloor+1 & \text{ if }\ \ \ \lfloor{N}\rfloor\ \text{is a
positive odd integer}\ ,
\end{cases}
\end{equation}
where
\begin{equation}\label{Eq49}
N({\bar a},l,m)\equiv {{ma}\over{\sqrt{a^2-M^2}}}-(l+1)\  .
\end{equation}
Interestingly, the fact that ultra-spinning (${\bar a}>1$) exotic
compact objects are characterized by a {\it finite} discrete family
$\{r_{\text{c}}({\bar a},l,m;n)\}^{n=N_{\text{r}}}_{n=1}$ of surface
radii that can support the static massless scalar field
configurations should be contrasted with the complementary case of
sub-critical spinning objects in the ${\bar a}<1$ regime which, as
previously proved in \cite{Pan,Hodeco2}, are characterized by an
{\it infinite} countable family $\{r_{\text{c}}({\bar
a},l,m;n)\}^{n=\infty}_{n=1}$ of surface radii that can support the
spatially regular static scalar fields \cite{Notefni}.

(5) The ability of {\it spinning} objects to support spatially
regular static scalar field configurations is physically intriguing
from the point of view of the no-hair theorems that have recently
been discussed in \cite{Hodrec,Hodrec2,Bha} for horizonless regular
spacetimes. In particular, it has been proved in
\cite{Hodrec,Hodrec2} that spherically-symmetric ({\it
non}-spinning) horizonless reflecting stars cannot support nonlinear
configurations made of massless scalar fields.

(6) It is important to stress the fact that, as shown in
\cite{Pan,Hodeco2}, the outermost (largest) surface radius
$r^{\text{max}}_{\text{c}}({\bar a})\equiv
\text{max}_n\{r_{\text{c}}({\bar a};n)\}$ that can support the
static scalar field configurations is of central physical importance
since it marks the boundary between stable
[$r_{\text{c}}>r^{\text{max}}_{\text{c}}({\bar a})$] and unstable
[$r_{\text{c}}<r^{\text{max}}_{\text{c}}({\bar a})$] composed
ultra-spinning-exotic-compact-object-massless-scalar-field
configurations.

(7) Solving numerically the analytically derived resonance equation
(\ref{Eq15}), we have demonstrated that the characteristic
supporting radius $r^{\text{max}}_{\text{c}}({\bar a},l,m)$ is a
monotonically decreasing function of the dimensionless rotation
parameter ${\bar a}$ of the ultra-spinning exotic compact objects
(see Table \ref{Table1}). Likewise, it has been demonstrated that
the critical (outermost) supporting surface radius
$r^{\text{max}}_{\text{c}}({\bar a},l,m)$ is a monotonically
increasing function of the harmonic parameter $l$ which
characterizes the massless scalar field modes (see Table
\ref{Table2}).

(8) We have explicitly shown that the characteristic resonance
equation (\ref{Eq15}) for the discrete family of supporting surface
radii is amenable to an analytical treatment in the physically
interesting regime $0<{\bar a}-1\ll1$ of {\it near-critical}
horizonless spinning objects. In particular, the remarkably compact
resonance formula [see Eqs. (\ref{Eq16}) and (\ref{Eq40})]
\begin{equation}\label{Eq50}
r_{\text{c}}(n)=M+{{2ma}\over{\pi(l+2n)}}\ \ \ ; \ \ \
n\in\mathbb{Z}\
\end{equation}
has been derived analytically for near-critical (${\bar a}\gtrsim
1$) composed
ultra-spinning-exotic-compact-object-massless-scalar-field
configurations in the $1\ll \pi(l+2n)\ll {{2m}/{\sqrt{1-{\bar
a}^{-2}}}}$ regime.

(9) We have verified that the predictions of the analytically
derived resonance formula (\ref{Eq50}), which determines the unique
family of surface radii of the near-critical ultra-spinning
($0<{\bar a}-1\ll1$) compact reflecting objects that can support the
static (marginally-stable) massless scalar field configurations,
agree remarkably well (see Table \ref{Table3}) with the
corresponding exact values of the supporting surface radii as
determined numerically from the characteristic resonance condition
(\ref{Eq15}).

(10) Finally, it has been proved that the resonance equation
(\ref{Eq15}) can be solved {\it analytically} in the physically
interesting case of ultra-spinning (${\bar a}>1$) exotic compact
objects whose reflecting surfaces coincide with the corresponding
horizon radius $r_{\text{c}}=M$ of extremal (${\bar a}=1$) Kerr
black holes with the same mass parameter. In particular, we have
derived the remarkably compact discrete resonance spectrum [see Eq.
(\ref{Eq43})]
\begin{equation}\label{Eq51}
{{a}\over{M}}={{1}\over{\sqrt{1-\big({{m}\over{l+2+2n}}\big)^2}}}\ \
\ ; \ \ \ n=0,1,2,...\
\end{equation}
for the ultra-spinning compact configurations with $r_{\text{c}}=M$
\cite{Noteast}. Interestingly, one finds from the analytically
derived resonance formula (\ref{Eq51}) that, in the $l+2n\gg m$
regime, the dimensionless angular momenta $\{{\bar
a}\}^{n=\infty}_{n=0}$ of the exotic ultra-spinning {\it reflecting}
objects with physical parameters $\{M,r_{\text{c}}=M\}$ can be made
arbitrarily close \cite{Notearc} to the corresponding dimensionless
angular momentum ${\bar a}_{EK}=1$ of an {\it absorbing} extremal
Kerr black hole with the {\it same} mass and radius.

\bigskip
\noindent
{\bf ACKNOWLEDGMENTS}
\bigskip

This research is supported by the Carmel Science Foundation. I thank
Yael Oren, Arbel M. Ongo, Ayelet B. Lata, and Alona B. Tea for
stimulating discussions.



\begin{thebibliography}{99}
\bibitem{Noteuns} It is important to note that it has recently been pointed out \cite{Hernw}
that exotic compact objects which respect the null energy condition
may be characterized by nonlinear instabilities to massless
perturbation fields which are related to the existence of stable
light rings in the corresponding horizonless spacetimes. Thus, in
order to be considered as physically acceptable quantum-gravity
alternatives to the canonical classical black-hole spacetimes,
ultra-compact exotic compact objects must violate the null energy
condition or alternatively have long instability time scales.
\bibitem{Noteun} We shall use natural units in which $G=c=\hbar=1$.
\bibitem{Notejm} Here $M$ and $J$ are respectively the mass and
angular momentum of the spinning exotic compact object.
\bibitem{Notemsm} It is worth noting that, while spinning exotic compact objects with reflecting
boundary conditions can support static (marginally-stable)
configurations of massless scalar fields \cite{Hodeco2}, these
spatially regular static fields cannot be supported by spinning Kerr
black holes with regular event horizons. Interestingly, however, it
has recently been revealed that stationary configurations made of
massive scalar fields can be supported in the exterior regions of
spinning black-hole spacetimes \cite{Hodrc,Herkr}.
\bibitem{Noteman} Following \cite{Pan,Hodeco2,PanCar},
we shall assume that the contributions of the compact reflective
surface to the energy and angular momentum of the curved spacetime
are negligible.
\bibitem{Notesk} As nicely emphasized in \cite{Pan,PanCar} (see also
\cite{kr1,kr2,kr3}), the Birkhoff theorem does not apply in
non-spherically symmetric spacetimes, and thus the exterior metric
of the spinning compact object is not unique. Following
\cite{Pan,Hodeco2,PanCar}, we shall assume that the external
spacetime of the horizonless exotic object is characterized by the
familiar metric components (\ref{Eq2}) of the spinning Kerr
spacetime.
\bibitem{Notedec} Here $\omega$ is the proper frequency of the massless
scalar field mode. The dimensionless parameters $\{l,m\}$ are the
harmonic (spheroidal and azimuthal) indices of the linearized field
with the mathematical property $l\geq |m|$.
\bibitem{Noteby} For brevity, we shall henceforth omit the dimensionless harmonic indices $\{l,m\}$.
\bibitem{Noteklm0} Here we have used the simple relation $K_{lm}(a\omega=0)=l(l+1)$ for
the static (marginally-stable) scalar field modes [see Eq.
(\ref{Eq6})].
\bibitem{Notexz} Here we have used the relation $x=({{z-i\sqrt{{\bar
a}^2-1}})/({z+i\sqrt{{\bar a}^2-1}}})$ [see Eqs. (\ref{Eq12}) and
(\ref{Eq16})].
\bibitem{Notesmm} It is worth pointing out that these $a\to -a$ and $m\to -m$
reflection symmetries also manifest themselves in the analytically
derived resonance equation (\ref{Eq19}) [Note that
${_2F_1}(a,b;c;z)\equiv{_2F_1}(b,a;c;z)$].
\bibitem{Notenao} We are not aware of any general proof in the
mathematical or physical literature for this interesting property of
the hypergeometric functions. However, we have verified numerically
the validity of the relation (\ref{Eq31}) for several values of the
arguments of the hypergeometric function
${_2F_1}[l+1-\gamma,l+1;2l+2;1-x_{\text{c}}(r)]$.
\bibitem{Notefp1} It is worth emphasizing that the asymptotic expansion
(\ref{Eq36}) of the hypergeometric function $_2F_1(a,b;c;z)$ is
valid in the regime $|b|\gg\text{max}\{|a|,|c|,|z|\}$ \cite{Itn}.
This strong inequality is satisfied in the near-critical regime
$\sqrt{1-{\bar a}^{-2}}\ll m/l$.
\bibitem{Notenn} Here we have used the relation $1=e^{-2i\pi n}$, where the integer
$n$ is the resonance parameter of the near-critical composed
ultra-spinning-exotic-compact-object-massless-scalar-field
configurations.
\bibitem{Notefp2} It is important to stress the fact that the asymptotic expansion
(\ref{Eq36}) of the hypergeometric function $_2F_1(a,b;c;y)$ also
requires the strong inequality $|by|\gg 1$ \cite{Itn}. If
$\pi(l+2n)\sqrt{1-{\bar a}^{-2}}/m\ll1$ then we find from Eqs.
(\ref{Eq13}), (\ref{Eq15}), and (\ref{Eq38}) the simple relation
$|by|=O[\pi(l+2n)]$ (note that $y\equiv 1-x_{\text{c}}$ in our
case), which implies that the asymptotic approximation (\ref{Eq36})
is valid in the regime $\pi(l+2n)\gg1$.
\bibitem{Noteapr} Here we have used the small-$x$ relation $\cot(x)=1/x+O(x)$.
\bibitem{Noteast} It is worth pointing out that the ultra-spinning exotic configurations
described by the resonance formula (\ref{Eq51}) with $n=0$ saturate
the bound (\ref{Eq45}).
\bibitem{Notearc} That is, ${\bar a}\to1^+$ in the asymptotic $(l+2n)/m\to\infty$ limit.
\end{thebibliography}
\end{document}